\documentclass[dvips]{article}
\usepackage{icrctc07}

\title{Results from the Blazar Monitoring Campaign at the Whipple 10m Gamma-ray Telescope}
\shorttitle{Results from the Blazar Monitoring Campaign at the Whipple 10m Gamma-ray Telescope}
\authors{David Steele$^{1}$ for the VERITAS Collaboration$^{9}$, and
         M. T. Carini$^{2}$,
         P. Charlot$^{3}$,
         O. Kurtanidze$^{4}$,
         A. Lahteenmaki$^{5}$,
         T. Montaruli$^{6}$,
         A. C. Sadun$^{7}$,
         M. Villata$^{8}$ }
\shortauthors{D. Steele and et al}
\afiliations{$^1$Adler Planetarium \& Astronomy Museum, 1300 S. Lakeshore Dr., Chicago, IL 60605, USA\\ 
$^2$Dept. of Physics and Astronomy, Western Kentucky University, 1906 College Heights Blvd \#11077, Bowling Green, KY 42101, USA \\
$^3$Laboratoire d'Astrophysique de Bordeaux, UMR 5804 LAB, 2 Rue de l'Observatoire, BP 89, 33270 Floirac, France \\
$^4$Abastumani Astrophysical Observatory, 383792 Abastumani, Georgia\\
$^5$Mets\"{a}hovi Radio Observatory, Mets\"{a}hovintie 114, FIN-02540 Kylm\"{a}l\"{a}, Finland\\
$^6$Dept. of Physics, Univ. of Wisconsin-Madison, 1150 University Ave, Madison, WI 53706, USA \\
$^7$University of Colorado at Denver, PO Box 173364, Denver, CO 80217, USA\\
$^8$Osservatorio Astronomico di Torino, via Osservatorio 20, 10025 Pino Torinese, Italy\\
$^9$For full author list see G. Maier, ``VERITAS: Status and Latest Results", 
these proceedings}
\email{contact: dsteele@adlerplanetarium.org}

\abstract{In September 2005, the observing 
program of the Whipple $10\,$m gamma-ray telescope was redefined to be dedicated almost exclusively to AGN 
monitoring. Since then the five Northern Hemisphere blazars that had already been detected 
at Whipple are monitored routinely each night that they are visible. Thanks to 
the efforts of a large number of multiwavelength collaborators, the first year of this program
has been very successful. We report here on the analysis of Markarian 421 observations taken from 
November, 2005 to May, 2006 in the gamma-ray, X-ray, optical and radio bands. }

\begin{document}
\maketitle

\section{Introduction}
Among active galactic nuclei (AGN), blazars are the most extreme and
powerful sources known and are believed to have their jets more
aligned with the line of sight than any other class of radio-loud AGN.
Blazars 
exhibit a broad continuum extending from the
radio through TeV $\gamma$-rays. 
and are characterized by large, rapid,
irregular amplitude variability in all accessible spectral bands. 
Their high level 
of variability makes long-term, well-sampled, multiwavelength observations 
of blazars very important for constraining and understanding their 
emission mechanisms and characteristic time-scales.

In September 2005, the observing program of the Whipple $10\,$m imaging 
atmospheric Cherenkov telescope (IACT) was 
redefined to be dedicated almost exclusively to nightly blazar monitoring. 
Since that time the five Northern Hemisphere blazars that had already been 
detected at Whipple (Mrk 421, H1426+428, Mrk 501, 
1ES 1959+650 and 1ES 2344+514) have been monitored routinely each night 
that they are visible above an elevation of 60$^\circ$. For 
the first time, this has provided the opportunity to obtain long-term and
well-sampled VHE light curves of these highly variable objects. Part
of the motivation for these observations was to provide a trigger for
more sensitive VHE observations of these AGN by the new generation of
IACT telescopes (CANGAROO-III, HESS, MAGIC and VERITAS) and to provide
baseline observations for similar observations with GLAST.

Here we will focus on the results of the monitoring campaign on 
Mrk 421. At a redshift of 
$z$=0.031, Mrk 421 was the first extragalactic source of VHE 
$\gamma$-rays to be discovered with ground-based telescopes \cite{a1}. Since then 
Mrk 421 has been studied extensively in the VHE regime
and many periods of intense variability
have been observed \cite{gaidos96, mazin2005, blazejowski2005}. 
Mrk 421 is a high-frequency-peaked BL Lac (HBL) object exhibiting a
non-thermal, broadband spectral energy distribution (SED) with two peaks, 
one in X-rays and a second in the TeV regime. The spectrum has been 
successfully reproduced by models where the X-ray peak is due to electron 
synchrotron radiation, and the TeV emission arises as the same electron 
population cools by inverse-Compton up-scattering of either the synchrotron 
photons (Synchrotron Self-Compton; SSC) or ambient photons (External 
Compton; EC) \cite{krawczynski2004}. Like all such objects studied to date, the two
peaks in the SED have been found to correlate
with each other both in peak frequency and in strength. 

The call for full multiwavelength coverage in previous
campaigns was only invoked when large flares were observed in the TeV
regime. For the campaign
reported in this work, we make no ``a priori" assumptions about flux
levels so that we can look for correlations on longer times scales
and under different source conditions.
Here we present light curves 
of radio, optical, X-ray and $\gamma$-ray data and a search for correlated 
emission between spectral bands.

\section{Methods}

\begin{figure*}
\begin{center}
\noindent
\includegraphics [width=0.99\textwidth]{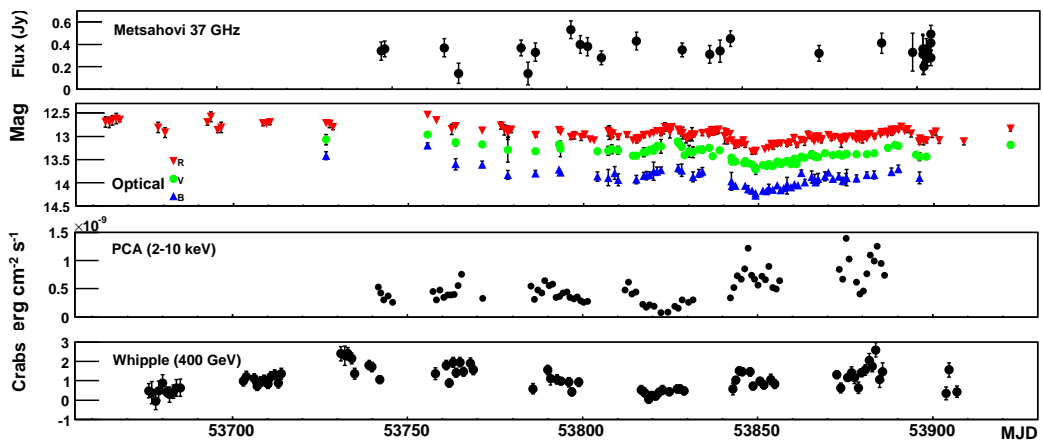}\\
\end{center}
\caption{Radio, optical, X-ray and $\gamma$-ray light curves of Mrk 421 in 2005-2006.}\label{fig:lightcurves}
\end{figure*}

All the TeV $\gamma$-ray observations presented here were made with the
$10\,$m Telescope at the Fred Lawrence Whipple Observatory (FLWO). 
During the period November 29, 2005 to May 29, 2006 a total of 98.0
hours of data were taken on Mrk 421. The data presented here were 
mostly taken in the {\it Tracking} mode 
and were analyzed using 
analysis procedures pioneered and developed by the Whipple 
Collaboration \cite{Horan07}.
Cuts were derived which
reject approximately 99.7\% of the cosmic-ray background images while retaining
over 50\% of those generated by $\gamma$-ray showers. 
A large sample of dark-field data (\mbox{$\approx$ 200 hours}) spanning a
similar zenith angle range was analyzed to estimate the background during
AGN data runs. 
Although sensitive in the energy range from $200\,$GeV to $10\,$TeV, the peak 
response energy of the telescope to a Crab-like spectrum during the 
observations reported upon here was approximately $400\,$GeV (subject
to a 20\% uncertainty). 

The X-ray data for this campaign were taken by 
the Rossi X-ray Timing Explorer (RXTE).
Data from 
the Proportional Counter Array (PCA) and the All Sky Monitor 
(ASM) were used, but only the PCA data are presented here. 
The sensitive energy range of PCA is 3 - 15 keV, limited
at the low-energy end by systematic effects in the detector, and at
the high-energy end by low source counts above the background \cite{johaod1996}.
Mrk 421 was observed with the RXTE PCA 
from January 6 to May 31, 2006 in short daily exposures of 19 minutes, on average.
Due to 
observing constraints,
the the X-ray and $\gamma$-ray observations were not always simultaneous,
with the PCA observations starting on average 99 minutes behind 
the Whipple observations. 
Analysis of the PCA observations followed the standard procedure outlined 
in the RXTE Cookbook.

Eight optical observatories contributed data sets resulting in excellent 
optical coverage over almost the entire length of the campaign.
They are the FLWO 1.2 m telescope; the Tenagra 0.8 m telescope in Tenagra, Arizona, USA; the
Bradford Robotic Telescope in Tenerife, Canary Islands, Spain; the WIYN
0.9 m telescope at Kitt Peak, Arizona, USA; the 0.7 m telescope at
Abastumani Observatory in Abastumani, Georgia; the 0.6 m Bell
Observatory at Western Kentucky University, Bowling Green, Kentucky, USA; Bordeaux Observatory
in Floirac, France; and the 1.05 m REOSC telescope at Osservatorio
Astronomico di Torino, Italy.  

The data from the various observatories were reduced and the
photometry performed independently by different analysts using
different strategies. 
Most analysts used relative aperture photometry performed 
using standard routines in IRAF \cite{iraf86}. Magnitudes were calculated with 
respect to one or more standard stars of
Villata \cite{Villata98}.
Because different observatories use different photometric systems, and 
photometric apertures and the definition of the reported measurement
errors for each nightly-averaged flux are inconsistent across
datasets, we have adopted a simple approach for the
construction of the composite light curves whereby a unique flux offset
is found for each spectral band of every instrument based on
nearly simultaneous observations between participating observatories. 


The radio observations of this campaign come from three observatories 
taken at 10 frequencies: Mets\"{a}hovi Radio Observatory, Kylmala, Finland 
(37 GHz), the University of Michigan Radio Astronomy Observatory, Dexter, 
Michigan, USA (4.8, 8.0 \& 14.5 GHz), and the Radio Astronomical Telescope of 
the Academy of Sciences (RATAN-600), Zelenchuksky, Russia (1, 2.3, 4.8, 8, 
11 \& 22 GHz). 
Only the 37 GHz observations are reported here; the 
full set of radio data, (as well as a full X-ray and $\gamma$-ray spectral 
variability analysis) will be presented elsewhere. 
The observations from Mets\"{a}hovi at 37 GHz have a bandwidth of 1 GHz and were 
obtained using the $14\,$m-diameter radio telescope. 

The resulting light curves for the spectral bands described above are 
presented in Fig. \ref{fig:lightcurves}. The source exhibits significant 
variability in all wavebands, with the level of variability generally higher
in the energy regimes corresponding to the two peaks of the SED. 

\section{Results \& Discussion}

\begin{figure}
\begin{center}
\noindent
\includegraphics [width=0.4\textwidth]{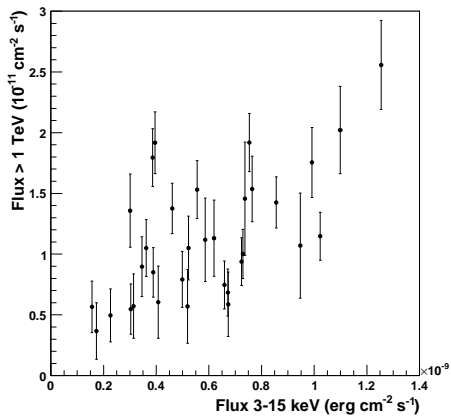}\\
\end{center}
\caption{TeV $\gamma$-ray flux vs. X-ray flux (3-15 keV).}\label{fig:x-gamma-flux}
\end{figure}

Of all the pairs of light curves investigated, the X-ray and TeV light curves
demonstrate the strongest correlation, as demonstrated by plotting the 
TeV flux vs. X-ray flux as in Fig. \ref{fig:x-gamma-flux}. The  
correlation suggests that the particle populations responsible
for at least part of the emission in the two energy regimes are collocated, 
as is the case in the SSC scenario. 

\begin{figure}
\begin{center}
\noindent
\includegraphics [width=0.48\textwidth]{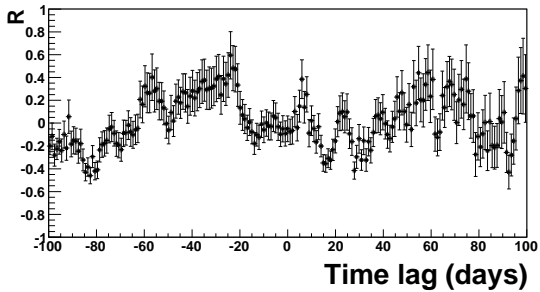}\\
\end{center}
\caption{Discrete Correlation Function between the optical R-band and
TeV band light curves. Positive values of time lag correspond to the optical
flux lagging the TeV flux.}\label{optTeVDCF}
\end{figure}

We searched for correlations between fluxes in other wavebands 
using the Discrete Correlation Function (DCF) \cite{dcf88}.  The DCF 
gives the linear correlation coefficient ($R$) for two light curves as a 
function of the time lag ($\tau$) between them.
Besides the X-ray and TeV light curves, only the DCF between the optical
and TeV light curves, presented in Fig. \ref{optTeVDCF}, warrants further 
investigation. The DCF indicates a possible correlation with the optical flux
lagging the TeV flux by about 7 days. An elevated level of
correlation is also seen with the optical leading the TeV by 25 to 60
days.

We investigated the significance of these possible correlations using
simulated optical light curves with the same variability properties as
the observed R-band light curve
based on a Power Spectral Density (PSD) derived from a fit to the first-order
structure function (SF). Assuming the
process(es) responsible for the optical variability are stationary
over the period of observation, the slope of the PSD, $\beta$, at a given 
variability frequency is related to the slope of the
SF, $\alpha$, by $\beta = 1 + \alpha$
\cite{hughes92}. 

After multiple realizations of the campaign using simulated, uncorrelated optical
light curves we assess the probability to have seen the observed
correlation features due to chance. For each bin of each simulated DCF, 
we record the values of the correlation coefficient, $R$, its error, $\sigma_R$, 
and $S = R /\sigma_{R}$. After many realizations, we build the
distribution of $S$ seen in each bin of the DCF and compute the chance
probability, $p_{i}$, to have found $S > S_{obs}$.
To assess the likelihood
of the observed DCF features as a whole, we construct two more
distributions, $\mathcal{L}_{\pm} = \prod_{i} p_{i}$ where for
$\mathcal{L}_{-}$ ($\mathcal{L}_{+}$) the product runs over all DCF bins 
with $60\,$d$ < \tau < 0$ ($0 \leq \tau < 60\,$d). From these, we find the likelihood to have
observed the optical precursor (optically-lagged) features by chance to be
$20\%$ ($60\%$).  Further investigation reveals that the lagged optical emission
feature seen at $\tau = 7$ d is likely due to contamination from the
R-band auto-correlation function, which also exhibits alternate peaks and valleys at
multiples of 7 days. The possible reality of the optical precursor
emission is more difficult to interpret since our method assesses all the
correlation features on a given side of the DCF at
once. Unfortunately, it would not have been possible to assess a
particular feature ``a posteriori" without an unknown trials penalty. 

\section{Conclusions}

We have presented multiwavelength observations of the blazar Mrk421 taken during
the first year of dedicated AGN monitoring with the Whipple $10\,$m Telescope.
The TeV $\gamma$-ray and X-ray fluxes were well-correlated, consistent with previous 
observations of objects of this class. 
There was 
an indication that the optical flux may have led the TeV flux by 25 to 60 
days, but a simulation reveals similar or stronger correlations arise by chance in one 
out of five simulated campaigns. 
We anticipate future long-term, multiwavelength 
observations will be important for further eludication of AGN emission mechanisms.

\section{Acknowledgments}
{\small
This research is supported by grants from the U.S. Department of Energy,
the U.S. National Science Foundation,
and the Smithsonian Institution, by NSERC in Canada, by PPARC in the UK and
by Science Foundation Ireland. This research uses data 
from the High Energy Astrophysics Science Archive Research Center (HEASARC), 
provided by NASA's Goddard Space Flight Center.
}

\bibliographystyle{plain}
{\small
\bibliography{icrc0728}
}

\end{document}